\documentstyle[prd,eqsecnum,aps]{revtex}

\newcommand{\bmit}[1]{\mbox{\boldmath $#1$}}
\newcommand{\calm}[1]{\frac{H_{#1}(a) - H_{#1}(a_0)}{\chi z}}
\newcommand{\calmp}[1]{\frac{h_{#1}(a) - h_{#1}(a_0)}{\chi z}}
\newcommand{\calmup}[2]{\frac{H_{#1}^{\ #2}(a) - H_{#1}^{\ #2}(a_0)}{\chi z}}
\newcommand{\calmpup}[2]{\frac{h_{#1}^{\ #2}(a) - h_{#1}^{\ #2}(a_0)}{\chi z}}

\newcommand{\calf}[1]{{\cal F}_{#1}}

\begin{document}
%\twocolumn
\draft
\preprint{Submitted to Classical and Quantum Gravity}
\title{On the propagation of electromagnetic radiation in the field of 
a plane gravitational wave}
\author{Enrico Montanari}
\address{Department of Physics, University of Ferrara and
INFN Sezione di Ferrara, Via Paradiso 12, I-44100 Ferrara, Italy}
\address{montanari@fe.infn.it}
\maketitle
\begin{abstract}
The propagation of free electromagnetic radiation in the field of a 
plane gravitational wave is investigated. A solution 
is found one order of approximation beyond the limit of geometrical 
optics in both transverse--traceless (TT) gauge and Fermi Normal 
Coordinate (FNC) system. 
The results are applied to the study of 
polarization perturbations. 
Two experimental schemes are investigated in order to verify the 
possibility to observe these perturbations, but it is found that the
effects are exceedingly small.   
\end{abstract}

%\narrowtext

%\twocolumn

\section{Introduction} 

The propagation of electromagnetic fields in curved space--time has 
always been an important problem since the formulation of general 
relativity.
Gravitational Doppler shift, deflection of light rays,
and gravitational lensing have been useful tests
for general relativity and have represented important tools in 
astronomical observations (see for instance~\cite{mtw,macho}).

In the first approach to this problem, the gravitational background 
was considered as a sort of anisotropic medium (a good 
introduction to this kind of approach is found in~\cite{str84}, while 
a recent exhaustive treatise is~\cite{gravlenses}; for the passage 
from wave optics to geometric optics in curved space--time 
see~\cite{ehl67}). In the literature, on the level of geometrical 
optics, many authors pointed 
their efforts at gravitational fields of isolated physical systems 
(such as a star at rest or a rotating body) which were not radiative
ones (see~\cite{bal58,ple60,hee64,pos65,muh66,fay82}). 
It was also considered the problem of light fluctuations in a 
gravitational wave background (see~\cite{zip66}); however in this work 
the author did not address his attention to possible polarization 
effects.

In the following years, in the framework of the linearized theory of 
general relativity, much work has been done to describe
the propagation of electromagnetic fields in a 
gravitational wave background, using a 
different approach with respect to the usual one. 
In fact many authors have split the electromagnetic tensor (or the
4--vector potential) in a sum of two terms: 
the first one is the flat space--time solution, while the second term 
describes the perturbation due to the weak gravitational field
(see~\cite{cooperstock68,boccaletti70,bertotti71,
funke78,codina80,baroni85,braginsky90,coo93}).
Nevertheless, as it will be shown in Sec. II, in this way the original 
physical problem is changed. 
It is for this reason that we think that the former approach is the 
most suitable for the problem we are concerned with.
An attempt in this direction has been recently done
(see~\cite{lobo92}) but
the Lorentz condition is not 
properly handled, being fulfilled only for particular directions of 
propagation and amplitudes of the 4--vector potential.

It is therefore seen that, as far as the problem of the propagation of 
electromagnetic radiation in the field of a gravitational wave is 
concerned, a satisfactory solution is lacking which takes into account 
the tensor nature of Maxwell equations, even in the geometrical 
optics limit.
Purpose of this paper is to investigate
this problem. Therefore, besides 
the usual phase perturbation, the consequences on the polarization of 
the electromagnetic field are evaluated.
It is a result of the present paper that, when Lorentz condition is 
met, nontrivial results in the polarization perturbations are found
(see Sec.~\ref{section7} and the conclusions).
 
As it is well known,
Maxwell equations are a set of redundant equations. If the problem is 
considered from the electromagnetic 4--potential point of view this 
redundance could be partially eliminated. 
The free electromagnetic equations in curved space time 
for the vector potential (de Rahm equations) read
\begin{equation}
A^{\mu ; \nu}_{\ \ \ \nu} - A^{\nu\ \ \mu}_{\ ;\nu} = 0
\label{derahm}
\end{equation}
(all through the paper, notations and conventions as in~\cite{mtw}).
In order to get a simple set of equations,
adopting the standard approach, we impose the Lorentz gauge condition 
on vector potential, bringing Eq.~(\ref{derahm}) into the system
\begin{equation}
\left \{ \begin{array}{rlc}
A^{\mu ; \nu}_{\ \ \ \nu} & = & 0 \\
A^{\mu}_{\ ;\mu} & = & 0
\end{array} \right.
\label{sistema1}
\end{equation}
Once a solution of the previous system is found, the antisymmetric 
tensor of the electromagnetic field, which is the physical measurable 
quantity, is obtained by the usual relation
\begin{equation}
F_{\mu\nu} = A_{\nu ;\mu} - A_{\mu ;\nu}
\label{fmunu}
\end{equation}
From the above equations it is clear that the direct interaction of a 
gravitational wave with an electromagnetic field is due only to the 
metric tensor. This means that in curved space--time the differential 
equations of electromagnetism have variable coefficients. 
However the explicit form of the dependence from the metric tensor 
depends critically on the reference frame used to perform calculation. 
Our strategy is to consider system~(\ref{sistema1}) in a reference 
frame in which calculation is easier. This system is the so called 
{\em transverse--traceless} (TT) gauge (see~\cite{mtw}). However the 
reference frame in which measures are performed is the {\em Fermi 
Normal Coordinate} (FNC) system (see for instance~\cite{mtw}). 
Therefore once the antisymmetric tensor of the electromagnetic field 
is known in the TT gauge we find its expression in FNC by means of the 
usual transformation rules (the electromagnetic field tensor is 
indicated by $F_{\mu\nu}$ in FNC and by ${\cal F}_{\alpha\beta}$ in TT 
gauge; $x^\mu$ are the FNC while $y^\alpha$ are the TT gauge ones):
\begin{equation}
F_{\mu\nu} = \frac{\partial y^\alpha}{\partial x^\mu}\,
\frac{\partial y^\beta}{\partial x^\nu}\, {\cal F}_{\alpha\beta}
\label{transfrule}
\end{equation}
The connection between the two frames is known for every point of the
space--time in the linear approximation of weak gravitational fields. 
This result is found in~\cite{fortini82,flores86}. 

The paper is organized as follows. In Sec.~\ref{section2} we find a 
solution to the de Rahm equations~(\ref{sistema1}) in TT gauge. In 
Sec.~\ref{section3} and Sec.~\ref{section4} we write the 
electromagnetic tensor components in TT gauge and FNC respectively.
Finally in Sec.~\ref{section7} we apply our results to the 
calculation of the polarization effects. 

\section{Solution of the de Rahm equations in TT gauge}
\label{section2}

For weak gravitational field, to which we shall limit ourselves in 
this paper, the metric tensor could be written as
\begin{equation}
g_{\mu\nu} = \eta_{\mu\nu} + h_{\mu\nu},\qquad 
|h_{\mu\nu}| << 1
\label{metric}
\end{equation}
where $\eta_{\mu\nu}$ is the Minkowski metric tensor with positive 
signature. In this approximation the particular coordinate frame 
called TT gauge (see~\cite{mtw}) is characterized by 
$h=\eta^{\mu\nu} h_{\mu\nu}=0$, $h_{0\mu}=0$ and $h^{ij}_{\ \ ,j}=0$.
Neglecting second order terms in $h_{\mu\nu}$ the linearized 
homogeneous de Rahm equations with the Lorentz condition are 
\begin{equation}
\left \{ \begin{array}{l}
A^{\mu,\nu}_{\ \ \,\nu} - h^{\alpha\beta} A^\mu_{\ ,\alpha,\beta} +
\left ( 
h^{\mu\ ,\beta}_{\ \alpha} + h^{\mu\beta}_{\ \ ,\alpha} -
h_\alpha^{\ \beta,\mu} \right ) A^\alpha_{\ ,\beta} = 0 \\
\ \\
A^\mu_{\ ,\mu} = 0
\end{array}
\right.
\label{system2}
\end{equation}
This system is a set of differential equations with variable 
coefficients. In the  
approach of~\cite{cooperstock68,boccaletti70,bertotti71,
funke78,codina80,baroni85,braginsky90} 
the vector potential $A^\mu$ is
split into two parts:
\begin{equation}
A^\mu = {^0\!A}^\mu + {^1\!A}^\mu
\label{split}
\end{equation}
where the first term is a solution of the flat space--time 
wave equation while the second one represents the perturbation due to 
the weak gravitational field. In this approach all the quadratic terms 
in the perturbation vector are neglected. With this assumptions 
system~(\ref{system2}) becomes a set of differential equations for the 
perturbed vector potential with constant coefficients (the d'Alembert 
operator) and a source term which involves 
the unperturbed potential and $h_{\mu\nu}$ (e.g.~\cite{codina80}). 
It is our opinion that
this procedure could in general lead 
to some misunderstanding on the nature of the interaction when the 
``unperturbed'' electromagnetic field is oscillatory (this is not the 
case of~\cite{boccaletti70} in which the ``unperturbed'' 
electromagnetic field is constant).
To show this point let us consider the case of an ordinary 
differential equation, the Mathieu equation (see for 
instance~\cite{mac,lm}), which may be written as
\begin{equation}
\ddot q + \omega^2 [1+h(t)] q = 0
\label{mathieu}
\end{equation}
and which describes, for instance, the charge on the plates of a 
condenser with a variable capacitance in an oscillating LC 
circuit~\cite{mac}. Supposing $h(t)<<1$, a splitting 
$q= {^0\!q} + {^1\!q}$ [in which $^0\! q$ is the ``unperturbed'' 
solution of Eq.~(\ref{mathieu}) when $h(t)=0$ and $^1\! q$ is the 
perturbation due to the presence of $h(t)$] brings to the equation of 
a harmonic oscillator for $^1\! q$ with a driven force proportional 
to the product of $h(t)$ and $^0\! q$. It is a matter of fact that, in 
general, the new set of equations {\em is not} equivalent to 
Eq.~(\ref{mathieu})~\cite{mac}. The physical reason
lies in the fact
that splitting of the charge changes the original problem of a free 
circuit with variable capacitance in a different problem of a circuit 
with constant capacitance and an external electromotive force.
In the same manner, splitting of the vector potential changes the 
problem of the propagation of a free electromagnetic field in a 
gravitational background, acting as a sort of anisotropic medium, in 
the problem of the generation of a field by a given current.
For these reasons our approach will be that of performing calculations 
{\em without} splitting of the vector potential. This strategy was yet 
carried out for the problem of the interaction of 
a gravitational wave with electromagnetic circuits 
in~\cite{fortini96}. 

Let us now state precisely the physical problem which we are intended 
to solve. Consider a TT gauge reference frame in which a plane
gravitational wave propagates in the $z$ direction. In this system the 
only non--vanishing components of $h_{\mu\nu}$ are 
[$\Phi_g=\chi\,(y^3-y^0)$]:
\begin{equation}
h_{11}(\Phi_g) = - h_{22}(\Phi_g) = h_+(\Phi_g); \qquad\qquad 
h_{12}(\Phi_g) = h_{21}(\Phi_g) = h_\times(\Phi_g)
\label{gravwave}
\end{equation}
Let us consider now a free electromagnetic field which in the absence 
of gravitational waves propagates 
in a given direction described by the spatial component of the four 
wave vector $k_\mu$.
The vector potential for such a field is given by 
$A^\mu = C^\mu g(k_\nu y^\nu)$, where $C^\mu$ are four constants
that satisfy Lorentz condition $C^\mu k_\mu=0$.
A gravitational wave, modifying the geometry of the 
space--time, acts as an anisotropic medium whose dielectric and 
magnetic properties change in space and in time, $1/\chi$ and 
$1/(c \chi)$ being the order of magnitude of space distances and 
time intervals for which variations take place appreciably. If
\begin{equation} 
|k_0| = k >>\chi,
\label{approx}
\end{equation}
then the amplitude, polarization and direction 
of the electromagnetic field remain practically constant over 
distances of the order of $1/k$ and times of the order of $1/(c k)$.
Therefore we can assume that 
the form of a solution of Eqs.~(\ref{system2}) would be
\begin{equation}
A^\mu = a^\mu f({\cal I})
\label{eikonal}
\end{equation}
where $a^\mu$ is a 4-vector function of space and time and ${\cal I}$ 
is a scalar function. In the framework of assumption~(\ref{approx}) we 
can suppose that first derivatives of $a^\mu$ and second derivatives of
${\cal I}$ are small quantities with respect to $a^\mu$ and $k_\mu$ 
respectively, so we can neglect superior order derivatives. 
This approach is one order of approximation beyond the geometrical 
optics limit allowing therefore to find not only the phase shift (that 
is to say the scalar function ${\cal I}$, the {\em eikonal}), but also 
the change in the polarization of the electromagnetic wave (for all 
these considerations see for instance~\cite{l2}). 

In order to solve the problem it is important to find, within the 
Lorentz condition, an electromagnetic gauge for which the equations 
are easier to solve. One can see immediately that, because of the 
space-time dependence of the gravitational wave 
[see Eq.~(\ref{gravwave})], the first equation of 
system~(\ref{system2}) is the 
same for $\mu=3,0$. Therefore we fix a gauge for which
\begin{equation}
A^3 = A^0
\label{newgauge}
\end{equation}
With this choice the equations for $A^r$ (from now on indices $r,s,t$ 
run from $1$ to $2$) are independent on $A^3=A^0$.
Once $A^r$ are found, by means of the Lorentz condition one can find 
immediately $A^3=A^0$ which fulfill also the first equation 
of~(\ref{system2}). To carrying on this program it is convenient to 
perform a coordinate change. We put
\begin{equation}
\left \{ \begin{array}{rcl}
T & = & (y^3-y^0)/\sqrt{2} \\
X & = & y^1 \\
Y & = & y^2 \\
Z & = & (y^3+y^0)/\sqrt{2}
\end{array} \right.
\label{uv}
\end{equation}
The new metric tensor $G_{\mu\nu}$ can be written as 
[see Eq.~(\ref{transfrule})]
\begin{equation}
G_{\mu\nu} = \frac{\partial y^\alpha}{\partial X^\mu}\,
\frac{\partial y^\beta}{\partial X^\nu}\, g_{\alpha\beta} = 
N_{\mu\nu} + h_{\mu\nu}(\sqrt{2} \chi T)
\label{newmetric}
\end{equation}
where
\begin{equation}
N_{\mu\nu} = \left (
\begin{array}{cccc}
0&0&0&1 \\
0&1&0&0 \\
0&0&1&0 \\
1&0&0&0 \end{array} \right )
\label{newflat}
\end{equation}
If we call ${\cal A}^\mu$ and $K^\mu$ the vector potential and the 
four wave vector in this new frame we get
\begin{equation}
{\cal A}^\mu = \left( \begin{array}{c}
(A^3 - A^0)/\sqrt{2} \\
A^1 \\
A^2 \\
(A^3 + A^0)/\sqrt{2} 
\end{array} \right ); \qquad
K_\mu = \left (
\frac{k_3 - k_0}{\sqrt{2}},k_1,k_2,\frac{k_3 + k_0}{\sqrt{2}}
\right )
\label{newvector}
\end{equation}
In this frame the fixed gauge is written as the usual Coulomb gauge in
flat space--time, that is to say:
\begin{equation}
{\cal A}^0 = 0; \qquad\qquad 
\frac{\partial {\cal A}^i}{\partial X^i} = 0.
\label{coulomb}
\end{equation}
The first equation of~(\ref{system2}) for the $t$ components is 
written as
\begin{equation}
N^{\alpha\beta} {\cal A}^t_{\ ,\alpha,\beta} - 
h^{rs} {\cal A}^t_{\ ,r,s} + h^t_{\ s,0} {\cal A}^s_{\ ,3} = 0
\label{teq}
\end{equation}
while the equation for ${\cal A}^3$ is given by
\begin{equation}
N^{\alpha\beta} {\cal A}^3_{\ ,\alpha,\beta} - 
h^{rs} {\cal A}^3_{\ ,r,s} - h_{r\ \ ,0}^{\ s} {\cal A}^r_{\ ,s} = 0
\label{3eq}
\end{equation}
Once ${\cal A}^t$ are found, from Eqs.~(\ref{coulomb}) we can find a 
general expression for ${\cal A}^3$; we can write
\begin{equation}
{\cal A}^3 = - \int^Z_0
\frac{\partial {\cal A}^r}{\partial X^r}\, dZ' + g(X,Y,T)
\label{a3coul}
\end{equation}
where, in order to fulfill Eq.~(\ref{3eq}), $g(X,Y,T)$ must be a 
solution of
\begin{equation}
\left [
\delta^{rs} - h^{rs}(\sqrt{2} \chi T) \right ] \,
\frac{\partial^2 g(X,Y,T)}{\partial X^r \partial X^s} = 
2 \left . \frac{\partial^2 {\cal A}^r}{\partial X^r\, \partial T}
\right |_{Z=0} +
h_{r\ ,0}^{\ s}(\sqrt{2} \chi T)
\left . \frac{\partial {\cal A}^r}{\partial X^s}
\right |_{Z=0}
\label{ga3}
\end{equation}
As we shall see later this function could be chosen in such a manner 
to have a solution which describes ``plane waves'' travelling in some 
direction.

In order to find a solution of Eq.~(\ref{teq}) we exploit consequences
of assumptions~(\ref{approx}) and~(\ref{eikonal}).
For what was said we neglect second order derivatives of $a^r$, and 
terms like $h_{\mu\nu,\alpha} a^r_{\ ,\sigma}$. Therefore 
Eq.~(\ref{teq}) is divided into the following equations for $a^\mu$ 
and ${\cal I}$:
\begin{eqnarray}
& G^{\alpha\beta} {\cal I}_{,\alpha} {\cal I}_{,\beta} = 0 & \\
& G^{\alpha\beta} \left (
2 a^t_{\ ,\alpha} {\cal I}_{,\beta} + a^t {\cal I}_{,\alpha,\beta}
\right ) +
\sqrt{2} \chi h^{(1)\,t}_{\ \ \ \,s} a^s {\cal I}_{,3} = 0 &
\label{teqai}
\end{eqnarray}
where $h^{(n)}_{rs}$ means the n--order derivative of $h_{rs}$ 
with respect to the argument $\sqrt{2} \chi T$. 

Now we notice that the ``unperturbed'' polarization amplitude and 
eikonal of the electromagnetic field are not oscillating quantity: 
indeed the polarization amplitude is a constant and the eikonal is a 
big quantity, changing by $2\,\pi$ when we move through one 
wavelength, while the gravitational perturbation occurs on the scale 
of $1/\chi >> 1/k$. Because of the smallness of the metric 
perturbation [see Eq.~(\ref{metric})] and of its coordinate 
dependence we can therefore safely put
\begin{eqnarray}
& {\cal I} = S + u(T) & \label{split2i} \\
& a^t = B^t + b^t(T)  &
\label{split2a}
\end{eqnarray}
where $S=k_\alpha x^\alpha + \phi$ is the usual phase of the flat 
space--time, $B^t$ are constant and finally $u(T)$ and $b^t(T)$ are 
the small perturbations due to the gravitational wave background.
With these assumptions Eq.~(\ref{teqai}) become
\begin{eqnarray}
\dot u & = & \frac{1}{2}\frac{K_r K_s}{K_3}\,h^{rs} \label{ufineq}\\
\dot b^r & = & - \frac{1}{2} \dot h^r_{\ s} B^s 
\label{bfineq}
\end{eqnarray}
where {\em dot} means derivation with respect to $T$.
The solutions to these equations are simply given by
\begin{eqnarray}
u & = & \frac{1}{2}\frac{K_r K_s}{\sqrt{2}\chi K_3}\,
\left [ H^{rs}(\sqrt{2}\chi T) - H^{rs}(\Phi_0)\right ] + u_0 
\label{uuvsol} \\
b^r & = & - \frac{1}{2} h^r_{\ s}(\sqrt{2}\chi T) B^s + b^r_0 
\label{buvsol}
\end{eqnarray}
where $H^{rs}(x)$ is a primitive of $h^{rs}(x)$.
As far as ${\cal A}^3$ is concerned, from Eqs.~(\ref{a3coul}), 
(\ref{split2i}), and~(\ref{split2a}) we can write
\begin{equation}
{\cal A}^3 = - \frac{K_r}{K_3} \left [
B^r + b^r(T) \right ] \left \{f[S + u(T)] - f[S_0 + u(T)] \right \} 
+ g(X,Y,T)
\label{uva3}
\end{equation}
where $S_0 = \left . S \right |_{Z=0}$.
Let us now fucus our attention on the function $g(X,Y,T)$. Taking into 
account Eqs.~(\ref{ufineq}), (\ref{bfineq}), and the relation
$N^{\alpha\beta} K_\alpha K_\beta =0$, then Eq.~(\ref{ga3}) could be 
written as
\begin{equation}
\left ( 
\frac{\partial^2\ }{\partial X^2} +
\frac{\partial^2\ }{\partial Y^2} -
h^{ts} \frac{\partial \ }{\partial X^t} 
\frac{\partial \ }{\partial X^s}
\right ) g =
- \frac{K_r}{K_3} \left (B^r + b^r \right ) 
\left (\delta^{ts} - h^{ts} \right ) K_t K_s f^{(2)}(S_0 + u)
\label{ga3sec}
\end{equation}
A solution of this equation is
\begin{equation}
g(T) = - \frac{K_r}{K_3} \left [B^r + b^r(T) \right ] f(S_0 + u)
\label{gsol}
\end{equation}
Putting this solution in Eq.~(\ref{uva3}) we find
\begin{equation}
{\cal A}^3 = - \frac{K_r}{K_3} \left [
B^r + b^r(T) \right ] f[S + u(T)]
\label{uva3f}
\end{equation}
The solution described by Eqs.~(\ref{coulomb}), 
(\ref{split2i})--(\ref{buvsol}), and (\ref{uva3f}) represents a 
``plane wave like'' solution for the electromagnetic field.

Getting back to the TT gauge system in which we wrote 
Eqs.~(\ref{system2}) we finally find 
\begin{eqnarray}
A^r & = & \left [
B^r - \frac{1}{2} h^r_{\ s}(\Phi_g) B^s + b^r_0 
\right ]
f \left (S + u \right ) \nonumber \\
& & \label{potsol} \\
A^0 = A^3 & = & - \frac{k_r}{k_3 + k_0} \left [
B^r - \frac{1}{2} h^r_{\ s}(\Phi_g) B^s + b^r_0 
\right ]\ f \left (S + u \right ) 
%+ v(x,y,z-\tau). 
\nonumber
\end{eqnarray}
where
\begin{equation}
u = \frac{1}{2} \frac{k_r k_s}{\chi (k_3 + k_0)} 
\left [
H^{rs}(\Phi_g) - H^{rs}(\Phi_0)
\right ] + u_0
\label{phase}
\end{equation}
is the phase shift function.
We have therefore written a plane wave like solutions of 
Eqs.~(\ref{system2}) in the particular Lorentz gauge for 
which $A^3=A^0$. As one can see from Eqs.~(\ref{potsol})
the amplitude of $A^3=A^0$ tends to infinity when $k_0 + k_3$ goes to 
zero (that is to say for parallel interaction). However this is only a 
result due to the particular chosen gauge, without any physical 
meaning. In fact in the next section we shall 
see that the electromagnetic field {\em does not} present such 
infinity problems.

\section{Electromagnetic field in TT gauge}
\label{section3}

In this section we find the components of the antisymmetric tensor of 
the electromagnetic field in interaction with a gravitational wave in 
the TT gauge. To this aim we shall use Eq.~(\ref{fmunu}). In order to 
do that we must before write the covariant component of $A^\mu$. We 
have
\begin{equation}
A_\nu = g_{\nu\mu} A^\mu \qquad \Rightarrow \qquad \left \{
\begin{array}{c}
A_3 = A^3 = A^0 = - A_0 \\
A_r = A^r + h_{rs} A^s
\end{array} \right.
\end{equation}
We therefore have:
\begin{eqnarray}
A_3 = - A_0 & = & - \frac{k_r}{k_3 + k_0} 
(B^r - \frac{1}{2} h^r_{\ s} B^s + b^r_0)\ f(S + u)
\nonumber \\
& & \label{covpot}\\
A_r & = & (B^r + \frac{1}{2} h^r_{\ s} B^s + b^r_0)\ f(S + u).
\nonumber 
\end{eqnarray}
In order to be consistent with the approximations made 
[see Eq.~(\ref{approx}) and its consequences] we neglect first order
derivatives of $h_{\mu\nu}$ in Eq.~(\ref{fmunu}).
The components of the electromagnetic field tensor in the TT gauge 
are therefore written as
\begin{eqnarray}
{\cal F}_{03} &=& - k_r \left (
B^r - \frac{1}{2} h^r_{\ s} B^s + b_0^r \right )\ f'(S+u) 
\nonumber \\
{\cal F}_{0r} &=& - \left [- k_0 \left (
B^r + \frac{1}{2} h^r_{\ s} B^s + b_0^r \right ) + u_{,3} B^r + 
\right. \label{electricfield} \\
& & \qquad\qquad\qquad\qquad\qquad\qquad
\left. + \frac{k_r k_s}{k_0 + k_3} \left (
B^s - \frac{1}{2} h^s_{\ t} B^t + b_0^s \right ) \right ]\ f'(S+u)
\nonumber 
\end{eqnarray}
and
\begin{eqnarray}
{\cal F}_{12} &=& \left [k_1 \left (
B^2 + \frac{1}{2} h^2_{\ s} B^s + b_0^2 \right ) -
k_2 \left (
B^1 + \frac{1}{2} h^1_{\ s} B^s + b_0^1 \right )
\right ]\ f'(S+u) \nonumber \\
{\cal F}_{r3} &=& - \left [k_3 \left (
B^r + \frac{1}{2} h^r_{\ s} B^s + b_0^r \right ) + u_{,3} B^r + 
\right. \label{magneticfield} \\
& & \qquad\qquad\qquad\qquad\qquad\qquad
\left. + \frac{k_r k_s}{k_0 + k_3} \left (
B^s - \frac{1}{2} h^s_{\ t} B^t + b_0^s \right ) \right ]\ f'(S+u)
\nonumber 
\end{eqnarray}
where we have used the fact that $u_{,0} = - u_{,3}$ and
\begin{equation}
u_{,3} =  \frac{1}{2} \frac{k_r k_s}{k_0 + k_3}\ 
h^{rs}(\Phi_g) 
\label{udz}
\end{equation}

\section{Electromagnetic field in FNC system}
\label{section4}

The aim of this section is to write the expression for the 
electromagnetic field tensor in FNC. In the framework of linear 
approximation, TT-coordinates are given by FNC by means of 
\begin{equation}
y^\mu(x^\alpha) = x^\mu + \epsilon^\mu(x^\beta)
\label{trule}
\end{equation}
which involves the infinitesimal vector field $\epsilon^\mu$;
this vector field is known in every point of the space 
(see~\cite{fortini82}) therefore solving the problem of the connection 
between FNC and TT--gauge. If the TT-gauge perturbation of the 
Minkowskian metric is a plane wave propagating in the $z$ direction, 
with angular frequency $\chi$ then the infinitesimal vector field
$\epsilon^\mu$ may be written as (see~\cite{flores86})
\begin{eqnarray}
\epsilon^v & = & \frac{1}{2 x^3} \left \{
\frac{H_{rs}[\chi(x^3-x^0)] - H_{rs}(-\chi x^0)}{\chi x^3} -
h_{rs}(-\chi x^0) \right \} x^r x^s \nonumber \\
 & & \label{flores} \\
\epsilon^r & = & - \left \{
\frac{H_{rs}[\chi(x^3-x^0)] - H_{rs}(-\chi x^0)}{\chi x^3} -
\frac{1}{2} h_{rs}(-\chi x^0) \right \} x^s \nonumber
\end{eqnarray}
where $v = 0,3$, and we remind that $H_{rs}(\phi)$ is a primitive of 
$h_{rs}(\phi)$ 
and that $r,s=1,2$.
Obviously a plane wave is only an approximation of the 
gravitational field in a relatively small region around a point far 
away from the source. This implies that~(\ref{flores}) represents the 
infinitesimal vector field $\epsilon^\mu$ only in the region in which 
the plane wave propagation is valid. In which follows we consider 
source distances and gravitational wavelengths for which 
Eqs.~(\ref{flores}) hold in the region of interest.

The electromagnetic tensor field components in FNC are given in terms 
of the TT gauge ones by means of Eq.~(\ref{transfrule}). If the 
transformation rule between the two system is given by 
Eq.~(\ref{trule}) then we have
\begin{equation}
F_{\mu\nu} = {\cal F}_{\mu\nu} + 
{\cal F}_{\mu\alpha} \epsilon^\alpha_{\ ,\nu} + 
{\cal F}_{\alpha\nu} \epsilon^\alpha_{\ ,\mu}
\label{ffnc1}
\end{equation}
In the range of validity of Eqs.~(\ref{flores}) for 
plane monochromatic waves from the $z$ direction we can write 
[$a=\chi(x^3-x^0)$, $a_0=-\chi x^0$]
\begin{eqnarray}
F_{0t} & = & \calf{0t} -
\left [ \calmup{t}{s} - \frac{1}{2} h_{t}^{\ s}(a_0) \right ] \calf{0s} + 
\nonumber \\
& + & \frac{x^s}{x^3} \left [ \calm{st} - h_{st}(a_0) \right ] 
\calf{03} + \label{0tovunque} \\
& - & \frac{\chi x^r x^s}{2 x^3} 
\left [ \calmp{rs} - h^{(1)}_{\ \ \,rs}(a_0) \right ] 
\left ( \calf{0t} - \calf{t3} \right ) + 
\nonumber \\
& + & \chi x^r \left [ \calmpup{r}{s} - 
\frac{1}{2} h^{(1)\ s}_{\ \ \,r}(a_0) \right ] \calf{st} \nonumber
\end{eqnarray}
\begin{eqnarray}
F_{03} & = & \calf{03} + \frac{x^r}{z} 
\left [ \calmup{r}{s} - h_r^{\ s}(a) \right ] \calf{0s} + 
\nonumber \\
& - & \frac{x^r x^s}{x^3 x^3} 
\left [ \calm{rs} - h_{rs}(a_0) - 
\frac{\chi x^3}{2} h^{(1)}_{\ \ \,rs}(a_0) \right ] \calf{03} +
\label{03ovunque} \\
& + & \chi x^r \left [ \calmpup{r}{s} -
\frac{1}{2} h^{(1)\ s}_{\ \ \,r}(a_0) \right ] \calf{s3}
\nonumber
\end{eqnarray}
\begin{equation}
F_{12} = \calf{12} + \frac{x^r}{x^3} \epsilon_{ts3} 
\left [ \calmup{r}{s} - h_{r}^{\ s}(a_0) \right ] \delta^{tj} 
\left ( \calf{j3} - \calf{0j} \right )
\label{12ovunque}
\end{equation}
\begin{eqnarray}
F_{t3} & = & \calf{t3} -
\left [ \calmup{t}{r} - \frac{1}{2} h_{t}^{\ r}(a_0) \right ] \calf{r3} +
\nonumber \\
& + & \frac{x^r}{x^3} 
\left [ \calm{rt} - h_{rt}(a_0) \right ] \calf{03} + \label{t3ovunque} \\
& + & \frac{x^r}{x^3}
\left [ \calmup{r}{s} - h_{r}^{\ s}(a) \right ] \calf{ts} + \nonumber \\
& + & \frac{x^r x^s}{x^3 x^3} 
\left [ \calm{rs} - \frac{1}{2} h_{rs}(a) - \frac{1}{2} h_{rs}(a_0)
\right ] 
\left ( \calf{0t} - \calf{t3} \right ) \nonumber
\end{eqnarray}
where $\epsilon_{ijk}$ is the completely antisymmetric unit 
pseudotensor whose components are zero unless 
$i\neq j\neq k$ and in this case they are equal to $1$ or $-1$ 
according to the fact that $ijk$ is a even or odd permutation of 
$123$.

If the electromagnetic field which is under investigation is localized 
in a space region with linear dimensions small compared with $1/\chi$ 
(for instance the electromagnetic field of a resonant cavity), then 
it is appropriate to expand the 
relations~(\ref{0tovunque})--(\ref{t3ovunque}) in power of 
$\chi x^i$. The first three terms of the expansion are given by:
\begin{eqnarray}
F_{0t} & = & {\cal F}_{0t} - \frac{1}{2} h_t^{\ r} {\cal F}_{0r} +
   \frac{\chi x^r}{2} \left ( h^{(1)\ s}_{\ \ \ r} {\cal F}_{st} +
         h^{(1)}_{\ \ \ rt} {\cal F}_{03} \right ) -
   \frac{\chi x^3}{2} h^{(1)\ r}_{\ \ \ t} {\cal F}_{0r} + 
\nonumber \\
& - & \frac{\chi^2 x^r x^s}{4} h^{(2)}_{\ \ \,rs} 
(\calf{0t} - \calf{t3}) + \label{ffnc0t} \\
& + & \frac{\chi^2 x^3 x^r}{2} \left ( \frac{1}{3} h^{(2)}_{\ \ \,rt}
\calf{03} + h^{(2)\ s}_{\ \ \,r} \calf{st} \right ) -
\frac{\chi^2 x^3 x^3}{6} h^{(2)\ s}_{\ \ \,t} \calf{0s} \nonumber
\end{eqnarray}
\begin{eqnarray}
F_{03} & = & {\cal F}_{03} - \frac{\chi x^r}{2} h^{(1)\ s}_{\ \ \ r} 
\left ( {\cal F}_{0s} - {\cal F}_{s3} \right ) + \nonumber \\
& - & \frac{\chi^2 x^r x^s}{6} h^{(2)}_{\ \ \,rs} \calf{03} -
\chi^2 x^r x^3 h^{(2)\ s}_{\ \ \,r} 
\left ( \frac{1}{3} \calf{0s} - \frac{1}{2} \calf{s3} \right )
\label{ffnc03}
\end{eqnarray}
\begin{eqnarray}
F_{12} & = & {\cal F}_{12} - \frac{\chi x^r}{2} h^{(1)\ t}_{\ \ \ r}
\epsilon_{st3} \delta^{sj} 
\left ( {\cal F}_{0s} - {\cal F}_{s3} \right ) + 
\nonumber \\
& + & \frac{\chi^2 x^r x^3}{6} \epsilon_{ts3} h^{(2)\ s}_{\ \ \,r}
\delta^{tj} (\calf{j3} - \calf{0j})
\label{ffnc12}
\end{eqnarray}
\begin{eqnarray}
F_{t3} & = & {\cal F}_{t3} - \frac{1}{2} h_t^{\ r} {\cal F}_{r3} + 
   \frac{\chi x^r}{2} \left ( 
    h^{(1)}_{\ \ \ rt} {\cal F}_{03} -
    h^{(1)\ s}_{\ \ \ r} {\cal F}_{ts} \right ) -
   \frac{\chi x^3}{2} h^{(1)\ r}_{\ \ \ t} {\cal F}_{r3} +
\nonumber \\
& - & \frac{\chi^2 x^r x^s}{12} h^{(2)}_{\ \ \,rs} 
(\calf{0t} - \calf{t3}) + \label{ffnct3} \\
& + & \frac{\chi^2 x^r x^3}{3} 
\left ( \frac{1}{2} h^{(2)}_{\ \ \,rt} \calf{03} - 
h^{(2)\ s}_{\ \ \,r} \calf{ts} \right ) -
\frac{\chi^2 x^3 x^3}{6} h^{(2)\ r}_{\ \ \,t} \calf{r3} \nonumber
\end{eqnarray}
in which explicit $h_{rs}$ and its derivatives are functions of 
$(-\chi x^0)$ and the tensor part of $\calf{\mu\nu}$ in the first 
term of~(\ref{ffnc0t})--(\ref{ffnct3}) is to be expanded in power of 
$\chi\,x^3$ until second order. We see immediately that the $z$ 
components of both electric and magnetic fields have the first term of
the expansion vanishing. Therefore in the first approximation the 
difference with the TT-gauge arises in the electromagnetic field 
component normal to the direction of propagation of the gravitational 
wave.

If the problem is to ``follow'' a photon in its motion, that is to say 
to consider the phase shift and the change in the polarization of a 
photon propagating in a gravitational wave field in a region small 
enough to use Eqs.~(\ref{ffnc0t})--(\ref{ffnct3}), then for consistency 
reasons we must also expand those equations in powers of $\chi x^0$. 
In this case we have:
\begin{eqnarray}
F_{0t} & = & {\cal F}_{0t} - \frac{1}{2} h_t^{\ r} {\cal F}_{0r} +
   \frac{\chi x^r}{2} \left ( h^{(1)\ s}_{\ \ \ r} {\cal F}_{st} +
         h^{(1)}_{\ \ \ rt} {\cal F}_{03} \right ) -
   \frac{\chi (x^3-x^0)}{2} h^{(1)\ r}_{\ \ \ t} {\cal F}_{0r} 
\nonumber \\
& - & \frac{\chi^2 x^r x^s}{4} h^{(2)}_{\ \ \,rs} 
(\calf{0t} - \calf{t3}) + \frac{\chi^2 x^r (x^3-x^0)}{2} 
h^{(2)\ s}_{\ \ \ r} {\cal F}_{st} + \label{ffnc0tt0} \\
& + & \frac{\chi^2 x^r (x^3-3 x^0)}{6}
h^{(2)}_{\ \ \ rt} {\cal F}_{03} -
\frac{\chi^2 (2 x^3 x^3 - 6 x^3 x^0 + 3 x^0 x^0)}{12}
h^{(2)\ r}_{\ \ \ t} {\cal F}_{0r} \nonumber
\end{eqnarray}
\begin{eqnarray}
F_{03} & = & {\cal F}_{03} - \frac{\chi x^r}{2} h^{(1)\ s}_{\ \ \ r} 
\left ( {\cal F}_{0s} - {\cal F}_{s3} \right )
- \frac{\chi^2 x^r x^s}{6} h^{(2)}_{\ \ \,rs} \calf{03} + 
\nonumber \\
& - & \frac{\chi^2 x^r (2 x^3 - 3 x^0)}{6} h^{(2)\ s}_{\ \ \ r}
{\cal F}_{0s} +
\frac{\chi^2 x^r (x^3 - x^0)}{2} h^{(2)\ s}_{\ \ \ r}
{\cal F}_{s3} \label{ffnc03t0}
\end{eqnarray}
\begin{eqnarray}
F_{12} & = & {\cal F}_{12} - \frac{\chi x^r}{2} h^{(1)\ t}_{\ \ \ r}
\epsilon_{st3} \delta^{sj} \left ( {\cal F}_{0j} - {\cal F}_{j3} \right )
+ \frac{\chi^2 x^r x^3}{6} \epsilon_{ts3} h^{(2)\ s}_{\ \ \,r}
\delta^{tj} (\calf{j3} - \calf{0j}) + \nonumber \\
& - & \frac{\chi^2 x^r (x^3-3 x^0)}{6} h^{(2)\ t}_{\ \ \,r}
\epsilon_{st3} \delta^{tj} (\calf{0j} - \calf{j3})
\label{ffnc12t0}
\end{eqnarray}
\begin{eqnarray}
F_{t3} & = & {\cal F}_{t3} - \frac{1}{2} h_t^{\ r} {\cal F}_{r3} + 
   \frac{\chi x^r}{2} \left ( 
    h^{(1)}_{\ \ \ rt} {\cal F}_{03} -
    h^{(1)\ s}_{\ \ \ r} {\cal F}_{ts} \right ) -
   \frac{\chi (x^3-x^0)}{2} h^{(1)\ r}_{\ \ \ t} {\cal F}_{r3} +
\nonumber \\
& - & \frac{\chi^2 x^r x^s}{12} h^{(2)}_{\ \ \,rs}
(\calf{0t} - \calf{t3}) -
\frac{\chi^2 x^r (2 x^3-3 x^0)}{6} h^{(2)\ s}_{\ \ \ r} {\cal F}_{ts} 
+ \label{ffnct3t0} \\
& + & \frac{\chi^2 x^r (x^3-3 x^0)}{6} h^{(2)}_{\ \ \ rt} 
{\cal F}_{03} -
\frac{\chi^2 (2 x^3 x^3 - 6 x^3 x^0 + 3 x^0 x^0)}{12}
h^{(1)\ r}_{\ \ \ t} {\cal F}_{r3} \nonumber
\end{eqnarray}
in which now explicit $h_{rs}$ and its derivatives are intended calculated 
in zero and also the tensor part of $\calf{\mu\nu}$ in the first 
term of~(\ref{ffnc0tt0})--(\ref{ffnct3t0}) is to be 
expanded in power of $\chi\,(x^3-x^0)$ until second order.

\section{Phase--shift}
\label{section5}

As a first application of the obtained results we calculate the 
total phase--shift between the light beams in an interferometer with 
arms in the directions defined by the azimuthal and polar angles 
$(\theta_r,\phi_r)$ ($r=1,2$). The 
gravitational wave propagates in the positive $z$ direction.
The total phase--shift is obtained by the difference of the phase 
change along the two arms:
\begin{equation}
\delta\Phi = \delta u_{1f} - \delta u_{2f} + 
\delta u_{1b} - \delta u_{2b}
\label{phaseshift}
\end{equation}
where lower indices $1,2$ refers to the two arms, and $f$, $b$ to 
forward and backward path.

The phase is a scalar quantity, therefore we can perform the 
calculation in TT gauge; the result holds true also in FNC, the 
laboratory frame.
In TT gauge, free falling bodies initially at rest, stay at rest at
any subsequent time (as it is easily seen by inspection of motion 
equations). Therefore, if the beam splitter is in the origin $O$ of 
the coordinates and the two mirrors are placed at a distance $L$ along 
the directions $\bmit n_r = (\sin{\theta_r} \cos{\phi_r}, 
\sin{\theta_r} \sin{\phi_r}, \cos{\theta_r})$ ($r=1,2$), the forward 
phase change is given by (see eq.~(\ref{phase})):
\begin{equation}
\delta u_{rf} = \frac{1}{2} \frac{k}{\chi} (1+\cos{\theta_r})
\left [H_{\phi_r}(\phi_0 - \chi y^0 - \chi L (1-\cos{\theta_r})) -
H_{\phi_r}(\phi_0 - \chi y^0) \right ]
\label{duf}
\end{equation}
where $H_\phi = \cos{2 \phi}\,H_+ + \sin{2 \phi}\,H_\times$.
When light is travelling backward the reverse direction of propagation 
is given by the angles $(\pi - \theta_r,\pi + \phi_r)$. Therefore the 
backward phase change is given by
\begin{equation}
\delta u_{rb} = \frac{1}{2} \frac{k}{\chi} (1-\cos{\theta_r})
\left [H_{\phi_r}(\phi_0 - \chi y^0 - 2 \chi L) -
H_{\phi_r}(\phi_0 - \chi y^0 - \chi L(1-\cos{\theta_r})) \right ]
\label{dub}
\end{equation}
This holds true for any waveform of the gravitational radiation. Now 
we consider a plane monochromatic gravitational wave. We set
\begin{equation}
h_\phi(\Phi_g) = A_\phi \sin{(\Phi_g+\alpha)};\quad\qquad\qquad 
A_\phi = \cos{2 \phi}\ A_+ + \sin{2 \phi}\ A_\times
\label{monochroma}
\end{equation}
where the constants $A_+$, $A_\times$ are the polarization amplitudes
($h_+(\Phi_g) = A_+\,\sin{(\Phi_g + \alpha)}$, 
$h_\times(\Phi_g) = A_\times\,\sin{(\Phi_g + \alpha)}$).
In this case we have
\begin{eqnarray}
\delta u_{rf} &=& A_{\phi_r} \frac{k}{\chi} (1+\cos{\theta_r})\ 
\sin{\left[\frac{\chi L}{2} (1-\cos{\theta_r})\right]}\  
\sin{\left [\frac{\chi L}{2} (1-\cos{\theta_r}) + \chi y^0 - \alpha
\right]}
\nonumber \\
\delta u_{rb} &=& A_{\phi_r} \frac{k}{\chi} (1-\cos{\theta_r})\ 
\sin{\chi L}\ \sin{(\chi L + \chi y^0 - \alpha)} + \nonumber \\
&-& A_{\phi_r} \frac{k}{\chi} (1-\cos{\theta_r})\ 
\sin{\left[\frac{\chi L}{2} (1-\cos{\theta_r})\right]}\  
\sin{\left [\frac{\chi L}{2} (1-\cos{\theta_r}) + \chi y^0 - \alpha
\right]} \nonumber 
\end{eqnarray}
Therefore the phase change in each arm is given by
\begin{eqnarray}
\delta u_{rf} + \delta u_{rb} &=& A_{\phi_r} \frac{k}{\chi}\ 
\left\{ 2 \cos{\theta_r}\ 
\sin{\left[\frac{\chi L}{2} (1-\cos{\theta_r})\right]}\  
\sin{\left [\frac{\chi L}{2} (1-\cos{\theta_r}) + \chi y^0 - \alpha
\right]} + \right. \nonumber \\
&+& \left. (1-\cos{\theta_r})\ 
\sin{\chi L}\ \sin{(\chi L + \chi y^0 - \alpha)} \right \}
\label{dur}
\end{eqnarray}
For small $\theta_r$ angle, when one arm direction is very near to 
that of $z$ axis one obtains
\begin{equation}
\delta u_{rf} + \delta u_{rb} 
\stackrel{\theta_r \rightarrow 0}{\longrightarrow}
A_{\phi_r} \frac{k}{\chi} \frac{\theta_r^2}{2} \left[
\sin{\chi L} \sin{(\chi L + \chi y^0 - \alpha)} +
\chi L \sin{(\chi y^0 - \alpha)} \right]
\label{dupar}
\end{equation}
while for almost antiparallel direction one has
\begin{equation}
\delta u_{rf} + \delta u_{rb} 
\stackrel{\theta_r \rightarrow \pi}{\longrightarrow}
A_{\phi_r} \frac{k}{\chi} \frac{(\pi-\theta_r)^2}{2} \left[
\sin{\chi L} \sin{(\chi L + \chi y^0 - \alpha)} +
\chi L \sin{(2 \chi L + \chi y^0 - \alpha)} \right]
\label{duantipar}
\end{equation}
We see therefore that for parallel and antiparallel directions 
there is no phase change.

In the limit of small distances $\chi L<<1$ one has
\begin{equation}
\delta u_{rf} + \delta u_{rb} 
\stackrel{\chi L << 1}{\longrightarrow}
A_{\phi_r}\,k\,L\,\sin^2{\theta}\,\sin{(\chi y^0 - \alpha)}
\label{shortphase}
\end{equation}

Now let us consider an interferometer with arms in the 
positive $x$ and $y$ directions respectively.
In this case $A_{\phi_1} = A_+$ and $A_{\phi_2} = - A_+$. Therefore
the total phase--shift is given by
\begin{equation}
\delta \Phi = 2 (\delta u_{1f} + \delta u_{1b}) =
A_+\,k\,(2 L)\ \frac{\sin{\chi L}}{\chi L}\ 
\sin{(\chi L + \chi y^0 - \alpha)}
\label{dphiperp}
\end{equation}
which is in agreement with known results (see for 
instance~\cite{saulson}).

%the mirrors at the end of 
%the arms are at a distance $L$ from the beam splitter, then 
%The phase being a scalar, the result is the same in any gauge 
%calculation performed in TT gauge 

\section{Polarization}
\label{section7}

There are two possible state of polarization. If the 
electromagnetic wave is propagating in the $\bmit k = k 
(\sin{\theta} \cos{\phi},\sin{\theta} \sin{\phi},\cos{\theta})$ 
direction we are naturally led to describe the electromagnetic field 
in a frame whose three axis directions are:
\begin{eqnarray}
\bmit e_{(1)} &=& (-\cos{\theta} \cos{\phi}, -\cos{\theta} \sin{\phi},
\sin{\theta})
\nonumber \\
\bmit e_{(2)} &=& (\sin{\phi}, -\cos{\phi},0)
\label{versors} \\
\bmit e_{(3)} &=& \frac{\bmit k}{k} =
(\sin{\theta} \cos{\phi},\sin{\theta} \sin{\phi},\cos{\theta})
\nonumber
\end{eqnarray}
In flat space--time, the possible polarization of the electric field 
are along the $\bmit e_{(1)}$ and $\bmit e_{(2)}$ directions. They are 
obtained when
\begin{eqnarray}
B_{(1)}^1 = -\cos{\phi}\,\frac{E_0}{k}; &\qquad\qquad&
B_{(1)}^2 = -\sin{\phi}\,\frac{E_0}{k}
\label{pol1}\\
B_{(2)}^1 = -\sin{\phi}\,\frac{E_0}{k}; &\qquad\qquad&
B_{(2)}^2 = \cos{\phi}\,\frac{E_0}{k}
\label{pol2}
\end{eqnarray}
When a polarized wave interacts with a gravitational wave, the 
electric field rotates and changes in magnitude. We can write
\begin{equation}
\left(\calf{k0}\right)_{(r)} = \left[(E_1)_{(r)} \bmit e_{(1)}^k +
(E_2)_{(r)} \bmit e_{(2)}^k + (E_3)_{(r)} \bmit e_{(3)}^k\right]
f'(S+u)
\label{decomposition}
\end{equation}
in which $r=1,2$ and
\begin{eqnarray}
(E_1)_{(1)} &=& E_0 \left[ 1 + \frac{1}{2} \left (
\cos^2{\theta} - \cos{\theta} -1 \right) h_\phi -
\frac{k}{E_0} \left(\cos{\phi} b_0^1 + \sin{\phi} b_0^2
\right)\right]
\nonumber \\
(E_2)_{(1)} &=& - E_0 \left[\frac{1}{2} h'_\phi + \frac{k}{E_0}
\left(\sin{\phi} b_0^1 - \cos{\phi} b_0^2\right)\right]
\label{decomp1} \\
(E_3)_{(1)} &=& \frac{E_0}{2} \sin{\theta} (1-\cos{\theta})
h_\phi \nonumber
\end{eqnarray}
\begin{eqnarray}
(E_1)_{(2)} &=& E_0 \left[\frac{1}{2} (1+2\cos{\theta}) h'_\phi 
- \frac{k}{E_0}
\left(\cos{\phi} b_0^1 + \sin{\phi} b_0^2\right)\right]
\nonumber \\
(E_2)_{(2)} &=& E_0 \left[ 1 - \frac{1}{2} \left (2 +
\cos{\theta}\right) h_\phi -
\frac{k}{E_0} \left(\sin{\phi} b_0^1 - \cos{\phi} b_0^2
\right)\right]
\label{decomp2} \\
(E_3)_{(2)} &=& - E_0 \sin{\theta} h'_\phi \nonumber
\end{eqnarray}
where
\begin{eqnarray}
h_\phi &=& \cos{2\phi}\ h_+ + \sin{2\phi}\ h_\times \\
h'_\phi &=& - \sin{2\phi}\ h_+ + \cos{2\phi}\ h_\times
\label{h'}
\end{eqnarray}

As an example of application, let us consider the following {\em 
``gedanken Experiment''}. A system is made by a 
polarizer in the origin of a TT reference frame and a particle with 
mass $m$ and charge $q$ at a distance $L$ along the direction 
$(\theta,\phi)$; both of them are freely falling and point--like with 
respect to the gravitational wavelength. 
An electromagnetic wave pulse is propagating in
the same direction. The duration $\tau_p$ of the pulse fulfills 
the condition $1/(c k) << \tau_p << 1/(c \chi)$, namely it is very 
small compared with the gravitational wave period but it is very long 
with respect to the electromagnetic period. In this way the 
electromagnetic pulse is short enough that the metric tensor could be 
considered constant in the region in which the electromagnetic field 
exists but also long enough that this could be considered practically 
a plane monochromatic wave. A charged particle in the presence of an 
electromagnetic field in curved space--time undergoes a force whose 
expression is
\begin{equation}
m c \left (
\frac{du^\mu}{ds} + \Gamma^\mu_{\,\alpha\beta} u^\alpha u^\beta
\right ) = \frac{q}{c} {\cal F}^\mu_{\ \nu} u^\nu
\label{lorentzgrav}
\end{equation}
where $ds$ is the line element, $u^\mu$ the four--velocity of 
the particle and $\Gamma^\mu_{\,\alpha\beta}$ are the 
Christoffel symbols.
If the particle is at rest in TT gauge, then $u^\mu = (1,0,0,0)$.
In this way one obtains
\begin{eqnarray}
\frac{d^2 y^0}{ds^2} &=& 0 \quad \Longrightarrow \quad
dy^0 = ds
\nonumber \\
\frac{d^2 y^3}{ds^2} &=& \frac{q}{m\,c^2}\ {\cal F}^3_{\ 0}
\label{lorentztt} \\
\frac{d^2 y^r}{ds^2} &=& \frac{q}{m\,c^2}\ {\cal F}^r_{\ 0}
\nonumber
\end{eqnarray} 
Let us suppose that the polarization is in the $\bmit{e}_{(2)}$ 
direction and that the electromagnetic beam reach the polarizer 
at the time $t_i$. Therefore at the particle position
\begin{equation}
{\cal F}^3_{\ 0} = - \frac{1}{2} E_0 \sin{\theta} 
\left[h'_\phi(a_f) - h'_\phi(a_i) \right] f'[k_\mu y^\mu + u(a_f)]
\label{f30pol}
\end{equation}
where $a_f=\chi(L \cos{\theta} - t_i - L)$,
$a_i=\chi(L \cos{\theta} - t_i)$ and 
$y^0 \in \left[t_i + L,t_i +L+\tau_p\right]$, 
$y^j \in \left[L\,e_{(3)}^i,(L+\tau_p)\,e_{(3)}^i\right]$.
The particle has therefore a TT acceleration in the direction $z$ 
given by
\begin{equation}
\frac{d^2y^3}{dt^2} = - \frac{q\,E_0}{2\,m} \sin{\theta}
\left[h'_\phi(a_f) - h'_\phi(a_i) \right] f'[k_\mu y^\mu + u(a_f)]
\label{ttacc3}
\end{equation}
This holds true for any waveform of the gravitational radiation. Now 
we consider a plane monochromatic gravitational wave [see 
eqs.~(\ref{monochroma})].
%We set
%\begin{equation}
%h_\phi(\Phi_g) = A_\phi \sin{(\Phi_g+\alpha)};\quad\qquad\qquad 
%A_\phi = \cos{2 \phi}\ A_+ + \sin{2 \phi}\ A_\times
%\label{monochroma}
%\end{equation}
%where the constants $A_+$, $A_\times$ are the polarization amplitudes
%[$h_+(\Phi_g) = A_+\,\sin{(\Phi_g + \alpha)}$, 
%$h_\times(\Phi_g) = A_\times\,\sin{(\Phi_g + \alpha)}$].
Therefore [see also Eq.~(\ref{h'})] 
$h'_\phi(\Phi_g) = A'_\phi \sin{(\Phi_g + \alpha)}$, 
$A'_\phi = - \sin{2\phi}\ A_+ + \cos{2\phi}\ A_\times$, and
\begin{equation}
\frac{d^2y^3}{dt^2} = - \frac{q\,E_0\,A'_\phi}{m} \sin{\theta}
\sin{\left[\frac{\chi\,L}{2} (1-\cos{\theta})\right]}\ 
\cos{\left[\alpha - \chi t_i -\frac{\chi L}{2} (1-\cos{\theta})
\right]}\ f'[k_\mu y^\mu + u(a_f)]
\label{ttacc3f}
\end{equation}
It is immediately seen that for parallel and antiparallel interaction 
the acceleration vanishes. For perpendicular interaction one has
\begin{equation}
\frac{d^2y^3}{dt^2} = - \frac{q\,E_0\,A'_\phi}{m} 
\sin{\left(\frac{\chi\,L}{2}\right)}\ 
\cos{\left(\chi t_i + \frac{\chi\,L}{2} - \alpha \right)}
f'[k_\mu y^\mu + u(a_f)]
\end{equation}
For $\chi L << 1$
\begin{equation}
\frac{d^2y^3}{dt^2} = - \frac{q\,E_0\,A'_\phi\,\chi\,L}{2\,m}
\sin{\theta} (1-\cos{\theta}) \cos{(\chi t_i - \alpha)}
f'[k_\mu y^\mu + u(a_f)]
\label{ttchil}
\end{equation}
from which one could easily see that the greater acceleration is 
caused when $\theta=2\,\pi/3$.
In order to get the observable acceleration, one must get its 
expression in FNC, being this the laboratory reference frame.
However the acceleration is a first order quantity in $h_{\mu\nu}$;
therefore its expression in FNC is the same as in TT gauge.

It has been shown that, in the laboratory reference frame, the 
particle is accelerated in the $z$ direction only because of the 
direct interaction between the incoming electromagnetic and 
gravitational waves. In the absence of one of the two wave the 
acceleration would vanish, thus being a peculiar feature of the 
electromagnetic propagation in a gravitational background.
This result contains an apparent paradox: the appearance of a 
component of the electric field along the gravitational wave 
propagation direction with the consequent 
acceleration of the particle along the wave vector of the 
gravitational wave, even though this last one is a transverse wave.
This is due to the fact that the components of the electromagnetic 
tensors are not independent (because of Maxwell equations).

In order to get the order of magnitude of the effect described by 
Eqs.~(\ref{ttacc3})--(\ref{ttchil}) let us suppose that the charged 
particle is a proton. Taking $\theta=\pi/2$, a laser beam with 
intensity $I=50\ {\rm W/mm^2}$ and wavelength 
$\lambda=2 \pi/k = 1\ {\rm \mu m}$, and putting the particle at a 
distance $L=\pi/\chi$ from the polarizer (in such a way to have the 
greatest acceleration) we find that, after the passage of the 
electromagnetic wave pulse, the drift velocity in the z direction is 
[the factor $\gamma$ takes into account the effect of the phases of 
the last two factors in Eq.~(\ref{ttacc3f})]
\begin{equation}
v_{d} = \gamma\ 0.7\ A'_\phi\ {\rm cm/sec},\qquad\qquad
-1\leq \gamma \leq 1;
\label{driftv}
\end{equation}
thus, even in this optimistic situation, the effect is exceedingly 
small and, most likely, out of the range of an experimental 
verification. 

As a second example of application let us consider the possibility to 
perform a polarization rotation measure. 
PVLAS is an experiment under construction which is designed to measure
the vacuum magnetic birefringence~\cite{pvlas}. The achievable 
sensitivity of about $5\ 10^{-9}\ {\rm rad/\sqrt{Hz}}$, with an 
integration time of $10^7\ {\rm sec}$~\cite{zavpc}, would allow the 
measurement of an ellipticity of $10^{-12}\ {\rm rad}$. With a 
simple modification of the experimental setup (the collocation of a 
quarter--wave plate) the same sensitivity could be reached in the 
rotation angle
of a linearly
polarized electromagnetic radiation~\cite{zavpc}. This is the best 
result to date. The description of the apparatus is found 
in~\cite{pvlas}; however, in the last years there have been many 
improvements in the performances. For instance the finesse of the 
$4.5\ {\rm m}$ Fabry--Perot cavity is expected to be 
$10^5 \div 10^6$, while 
the laser have an intensity of $\sim 100\ {\rm mW/mm^2}$~\cite{zavpc}.
In the same assumptions as in the previous example for the 
electromagnetic pulse, and for gravitational waves of 
$\sim 100\ {\rm Hz}$, the approximation $\chi L \ll 1$ could be used, 
and then Eq.~(\ref{f30pol}) gives on the analyzer prism
\begin{equation}
F_{03} = {\cal F}_{03} =  
F\,\frac{\chi\,L}{2}\,E_0\,A'_\phi\,\sin{\theta}\,
\cos{(\chi t_i - \alpha)} f'[k_\mu y^\mu + u(a_f)]
\label{f03pvlas}
\end{equation}
where $F$ is the finesse of the cavity and the first equality holds 
for first order quantities in $h_{\mu\nu}$. The rotation 
angle is therefore given by 
\begin{equation}
\sin{\theta}\,F\,\frac{\chi\,L}{2}\,A'_\phi \sim 
\sin{\theta}\ A'_\phi.
\label{anglepvlas}
\end{equation}
It is therefore seen once again that also 
taking the most optimistic value of $10^{-24}$ for $A'_\phi$ 
(gravitational wave from a pulsar) the effect is too small to be 
detected.

By the way we note that in Ref.~\cite{coo93} authors claimed that no 
rotation of the plane of the polarization of the electromagnetic wave 
occurs to first order in $h$ calculations. Indeed this result was 
obtained for a particular linear polarization of the gravitational 
wave. In fact authors set $h_\times=0$. In this particular case also 
our solution is consistent with the previous claim. However in the 
general case there is a rotation of the plane of the polarization of 
the electromagnetic wave. 
%%It is interesting to note that also in the 
%%framework of their approach this would be the case.
%We also point out that when the gravitational frequency becomes too 
%small with respect the electromagnetic one, the solution obtained by 
%the splitting approach 
%diverges 
%becomes inconsistent with the (to this purpose see for 
%instance~\cite{cal98}).
%. A closer investigation to the 
%physical situation considered in Ref.~\cite{coo93}, with the approach 
%of the present paper can be found in~\cite{cal98}.
A comparison showing more in detail the differences between the 
approach of this paper and the splitting procedure in the physical 
situation described in~\cite{coo93} is found in~\cite{cal98}.

\section{Conclusion}
\label{conclusion}

We have found a solution in TT gauge to the free de Rahm equations 
with Lorentz condition for a gravitational background described by 
a plane wave, within the framework of two 
approximations. The first one is the linear gravitational 
approximation. The second one is that the gravitational wavelength 
is much smaller than the electromagnetic one; this allowed us to find 
a solution one order of approximation beyond the limit of geometrical 
optics. 
The result was used to write the components of the electromagnetic 
tensor in TT gauge and FNC which automatically fulfill free Maxwell 
equations in curved space--time up to terms of order $(\chi/k)^0$. 

We have applied these solutions to the problem of 
polarization of an electromagnetic field, 
exploiting an interesting feature of the propagation of 
electromagnetic waves in a gravitational wave background. In fact we 
have found that,
when an electromagnetic field linearly polarized interacts with
a gravitational wave whose direction of propagation is perpendicular 
to the electric field, then the electromagnetic field changes in 
magnitude and rotates in such a way that the electric field gets a 
component parallel to the gravitational wave vector. This result could 
seem in disagreement with the transversal nature of the gravitational 
waves: however this fact arise from the mutual dependence of the 
electromagnetic tensor components through Maxwell equations.

Two possible applications of 
this result have been investigated. 
However the effects are so small that are out of the range of 
current experimental techniques, and unlikely to be ever observed 
in the future. 
We think that
a study is recommended in order to 
check if possible measurable effects could be found on light beam 
coming from cosmological distances
(analogously to what was done 
in~\cite{zip66} for amplitude fluctuations).

\acknowledgments

The author wish to thank M. Calura, A. Sabbioni, P. Fortini, and 
C. Gualdi for illuminating discussions. The author wish also to thank 
G. Zavattini for his kind explanation of the PVLAS experiment and D. 
Etro for invaluable help.

\end{document}